\documentclass[12pt,preprint]{elsart}

\usepackage{latexsym}
\usepackage{amsmath}
\usepackage{amsfonts} 
\usepackage{epsfig}

\newcommand{\abs}[1]{\left| #1 \right|}

\newcommand{\bea}{\begin{eqnarray}}
\newcommand{\eea}{\end{eqnarray}}
\newcommand{\beann}{\begin{eqnarray*}}
\newcommand{\eeann}{\end{eqnarray*}}

\newcommand{\order}[1]{{\mathcal O}\left( #1 \right)}

\newcommand{\nn}{\nonumber}

\begin{document} 

\begin{frontmatter}
\title{Kinetic and Dynamic Delaunay tetrahedralizations in three dimensions}
\author{Gernot Schaller\thanksref{mail}}
\author{and Michael Meyer-Hermann}
\address{Institut f\"ur Theoretische Physik,
	Technische  Universit\"at Dresden,
	D-01062  Dresden,
	Germany}
\thanks[mail]{Corresponding author, electronic address :
{\tt schaller@theory.phy.tu-dresden.de}}
\date{\today}

\begin{abstract} 
We describe the implementation of algorithms to construct and maintain three-dimensional 
dynamic Delaunay triangulations with kinetic vertices using a three-simplex data structure. 
The code is capable of constructing the geometric dual, the 
Voronoi or Dirichlet tessellation.
Initially, a given list of points is triangulated.
Time evolution of the triangulation is not only governed by kinetic vertices but also
by a changing number of vertices.
We use three-dimensional simplex flip algorithms, a stochastic visibility walk algorithm for 
point location and in addition, we propose a new simple method of 
deleting vertices from an existing three-dimensional Delaunay 
triangulation while maintaining the Delaunay property.
The dual Dirichlet tessellation can be used to solve differential 
equations on an irregular grid, to define partitions in cell tissue
simulations, for collision detection etc.
\end{abstract}    


{\rm\small Pacs: 02.40.S, 02.10.R}

\end{frontmatter}

In nearly all aspects of science nowadays simulations of discrete objects 
underlying different interactions play a very important role. Such an 
interaction for example could be mediated by colliding grains of sand in an 
hourglass or -- more abstract -- the neighborhood question of
influence regions.
One general method to represent possible two-body interactions within a 
system of $N$ objects is given by a network which can be described by an 
$N \times N$ adjacency matrix $\nu$, with its matrix elements 
$\nu_{ij}=\nu_{ji}$ (undirected graph) representing the interaction between 
the objects $i$ and $j$.
However, for most realistic systems the graph defined this way is not 
practical if one remembers that the typical size of a system of atoms in 
chemistry can be $\order{10^{23}}$, the human body consists of 
$\order{10^{18}}$ cells and even simple systems such as a grain-filled 
hourglass contain $\order{10^{3}}$ constituents.
Even some reasonable fraction of such systems would be far too complex to be 
simulated by adjacency matrices. However, in most systems in reality the 
interactions at work have only a limited range. Physical contact forces such as
adhesion for example, can only be mediated between next neighbors. 
In such systems the adjacency matrix elements $\nu_{ij}$ would vanish for all 
distant constituents and therefore a more efficient description can be given by a sparse 
graph, where the adjacency relations between objects moving in their parameter 
space can be updated using rather simple methods.
In some systems -- such as solids crystallizing in a lattice configuration -- 
the neighborship are uniform and therefore a priori known. This can be effectively 
exploited if one considers all deviations from a lattice configuration as small 
perturbations. 
But also rather heterogenous systems can be modeled by lattice methods, 
e.g. the method of cellular automata \cite{neumann1966} has been used to model
cell tissues \cite{baer1974,meyerhermann2002}.
Note however, that several adaptations have to be performed in order to account 
to the different nature of next-neighborship in these systems.
In the realistic system of cells in a human tissue for example, the number of
next neighbors per cell is neither constant over the cell ensemble nor are the
interaction forces. To make things worse, all these parameters become time
dependent for dynamic systems.
The same holds true in the framework of collision detection.

In this article it is our aim to describe our implementation of a code 
generating three-dimensional dynamic (supporting insertion and deletion) and 
kinetic (supporting point movements) Delaunay triangulations. 
Delaunay triangulations and their geometric dual -- the Voronoi tessellation -- 
have been demonstrated before to be suitable tools to model cell tissues 
\cite{Meineke2001,Weliky1990,Weliky1991}.
However, these considerations have been restricted to the two-dimensional case.
In three-dimensional space kinetic (but not fully dynamic) Delaunay triangulations 
have been applied e.g. in the framework of collision detection amongst spherical grains
\cite{Ferrez2001}.

The generation of Delaunay triangulations is a 
well-covered topic, for a review see e.g.~\cite{okabe2000,berg1997,Fortune1992}.
Such triangulations are widely used for grid generation in finite element 
calculations and surface generation for image analysis \cite{luerig1996}. 
Since the Delaunay triangulation in general tries to avoid flat simplices, 
it also produces a good quality mesh for the solution of differential 
equations \cite{Miller1995,Bottino2000}.
In dimensions higher than two however, the situation is much more complicated. For example,
three-dimensional triangulations of the same number of points may have 
different numbers of tetrahedra \cite{okabe2000}. This can be compensated by
using more dynamic data structures that allow for a varying number of 
simplices such as lists. A more serious problem however, is posed
by the fact, that a two-dimensional polygon can always be triangulated,
whereas a three-dimensional non-convex polyhedron may not admit a decomposition
in tetrahedra without using artificial (Steiner) points 
\cite{schoenhardt1928,Goodman1997}. 
These differences result in the important consequence that not all algorithms 
can be generalized in a straightforward way from two-dimensional Delaunay
triangulations.
The maintenance of the triangulation in the case of dynamic (moving) vertices
now requires a data structure capable of handling a varying number of 
simplices in time.
Another important problem is the deletion of vertices from a Delaunay 
triangulation which is simple in two dimensions 
\cite{Brouns2001,Devillers2002} but transforms into a nontrivial problem
in higher dimensions \cite{shewchuk2002}, because in three dimensions there may
exist non-convex polyhedra (e.g.~Sch\"onhardt's polyhedron \cite{schoenhardt1928}) 
that cannot be tetrahedralized \cite{shewchuk1998}.\\
We have implemented algorithms for both adding and deleting vertices to 
a three-dimensional Delaunay triangulation that are incremental in the sense 
that they transform a Delaunay triangulation with $n$ vertices into a Delaunay
triangulation with $(n+1)$ or $(n-1)$ vertices, respectively. 
In addition, we have implemented flip algorithms 
\cite{Ferrez2001,Edelsbrunner1996,muecke1998} to maintain the Delaunay-property
of the triangulation in the case of kinetic vertices.
These ingredients together allow to provide a code for dynamic three-dimensional
Delaunay triangulations with kinetic vertices.
Such a code is suitable for the construction and maintenance of proximity structures
for moving objects, e.g. cell tissue simulations, where cell 
proliferation, cell death, and cell movement are essential elements that have not been covered
by Delaunay triangulations in three dimensions before.
Since for many neighborhood interaction forces (especially in cell tissues),
the contact surfaces and volumes of the dual Dirichlet tessellation is of 
importance, we have also implemented algorithms to compute these values from 
a given Delaunay triangulation.

This article is organized as follows:\\
In section \ref{Sdelaunay} we briefly review the concept of the Delaunay
triangulation by addressing the basic conventions in \ref{SSconventions}, the
elementary topological transformations in a triangulation in \ref{SSett}, 
defining the Delaunay criterion in \ref{SSdelaunay} and considering the geometric
dual in \ref{SSvoronoi} as well as the more technical volume and surface 
calculation of Voronoi cells in \ref{SSsurface_volume}.
In section \ref{Simplementation} we describe the actual implementation of the
algorithms by describing the used data structure in \ref{SSdatastructure}, an
incremental insertion algorithm in \ref{SSincinsert}, a stochastic visibility walk
algorithm in \ref{SSlocation}, the used flip algorithms in \ref{SSflips}, and
close with a description of an algorithm for incremental vertex deletion in
\ref{SSdeletion}.
In section \ref{Sresults} we analyze performances of the incremental insertion
algorithm in \ref{SSincinsalg}, the incremental deletion algorithm in 
\ref{SSincdeletalg}, the transformation of slightly perturbed Delaunay
triangulations into Delaunay triangulations in \ref{SSrestore}, and finally 
consider the performance of a combination of all processes in \ref{SSmixed}.
Robustness ist also briefly addressed.
We will close with a summary in section \ref{Ssummary}.

\section{The Delaunay Triangulation}\label{Sdelaunay}

\subsection{Conventions}\label{SSconventions}

For the sake of clarity, the illustrations in this article will be 
two-dimensional, unless noted otherwise.
Following the notation in the literature \cite{Edelsbrunner1996,muecke1998} we 
denote by the term vertex a position\footnote{If one extends the algorithms
towards weighted triangulations a vertex in addition contains a 
weight.} in three-dimensional space. By an {\it $n$-simplex} in 
${\mathbb R}^d$ 
($n\le d$) we understand the convex hull of a set $T$ of $n+1$ affinely 
independent vertices, which reduces in the three-dimensional case to 
tetrahedra 
(3-simplices), triangles (2-simplices), edges (1-simplices) and vertices 
(0-simplices). Every $n$-simplex has a uniquely defined $n$-circumsphere.
Recall that a tetrahedron is bound by four triangles, six edges and four 
points in three dimensions.
These $(n \le d)$-simplices $\sigma_U$ -- formed by the convex hull of a 
subset $U \subseteq T$ -- are also called faces of $T$.
Since we will work in three dimensions,  we will shortly denote 3-simplices by
the term simplex.
A collection of these simplices $\mathcal K$ is called a 
{\it simplicial complex} if:
\begin{itemize}
\item{The faces of every simplex in $\mathcal K$ are also in $\mathcal K$}
\item{If $\sigma_T \in \mathcal K$ and $\sigma_{T^\prime} \in \mathcal K$,
	then $\sigma_T \cap \sigma_{T^\prime}=\sigma_{T \cap T^\prime}$.\\
	(the intersection of two simplices is at most a face of both, the
	simplices are 'disjoint')}
\end{itemize}
In numerical calculations with kinetic vertices the above criterion can be
destroyed: A vertex might move inside another simplex thus yielding two 
$n$-simplices whose intersection is again an $n$-simplex. 
We will refer to this situation as an invalid triangulation.
To be more exact, a triangulation is defined as follows.
If $S$ is a finite set of points in ${\mathbb R}^d$, then a simplicial 
complex $\mathcal K$ is called a {\it triangulation} of $S$ if
\begin{itemize}
\item{each vertex of $\mathcal K$ is in $S$}
\item{the underlying space of $\mathcal K$ is $conv(S)$}
\end{itemize}
By the degree of a vertex in a triangulation we will denote the number of
simplices in the triangulation containing the vertex as endpoints.
Furthermore, we will use the terms tetrahedralization and triangulation in
three dimensions synonymously, unless noted otherwise.

\subsection{Elementary Topological Transformations}\label{SSett}

\begin{figure}[ht]
  \centering {\includegraphics[height=6cm]{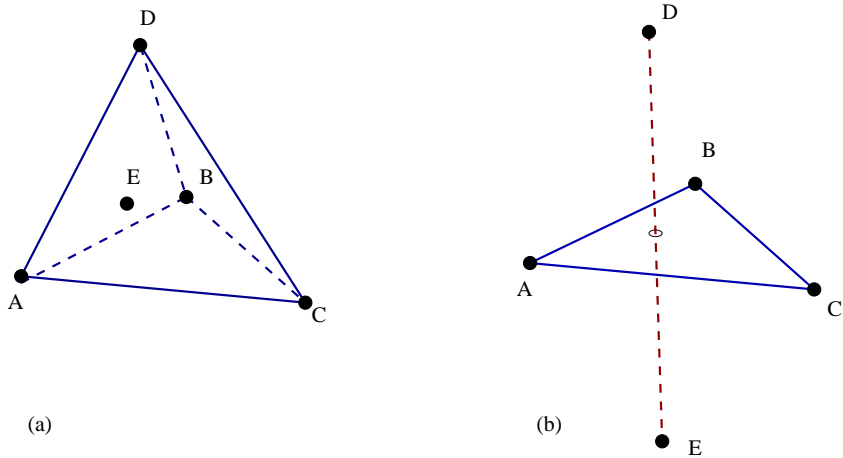}}
  \caption{\label{Fradon}
	Illustration of the Radon partition in three dimensions. There are 
	two possible constellations of the 5 points A, B, C, D, E in three
	dimensions. In (a) the point E lies within the simplex 
	formed by (A, B, C, D), whereas in (b) none of the
	vertices lies within the simplex formed by the other ones.}
\end{figure}

To an existing triangulation in ${\mathbb R}^3$ several topological
transformations can be applied. We will briefly remind the 
main ideas. For a more detailed discussion see e.g.~
\cite{Edelsbrunner1996,Xinjian1997}.
The discussion basically relies on Radon's theorem 
(see e.g.~\cite{Goodman1997,Edelsbrunner1996}):\\
\\
Let $X$ be a set of $d+2$ points in ${\mathbb R}^d$. Then a partition 
$X=X_1 \cup X_2$ with $X_1 \cap X_2 = \emptyset$ exists such that 
$conv(X_1) \cup conv(X_2) \neq \emptyset$.\\
If $X$ is in general position -- meaning that every subset of $X$ with at most
$d+1$ elements is affinely independent -- then this partition is also unique.
In our case this simply means that \cite{Fortune1992,muecke1998}
\begin{itemize}
\item{no four points lie on a common plane}
\item{no five points lie on a common sphere}
\end{itemize}
Figure \ref{Fradon} illustrates the idea of Radon's theorem in three 
dimensions.

\begin{figure}[ht]
  \centering {\includegraphics[height=6cm]{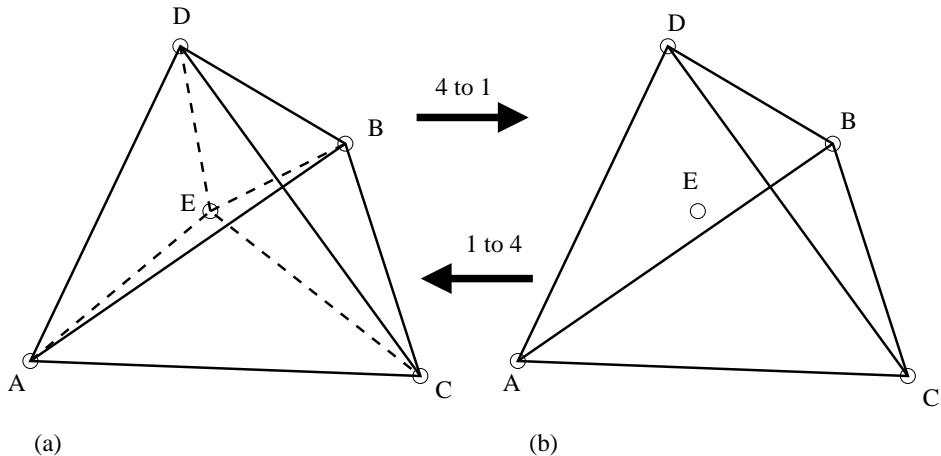}}
  \caption{\label{Fflip14}
	Three-dimensional illustration of the addition or deletion of a
	vertex E. In the left picture (a) one has exactly four simplices: 
	(A, B, C, E), (A, B, D, E), (A, C, D, E), (B, C, D, E), whereas in the
	right picture (b) the vertex E lies within the simplex (A, B, C, D). 
	The edges that can principally not be seen from the outside are
	drawn dashed.
	Switching between the two configuration corresponds to adding 
	($1 \to 4$) the vertex E to an existing triangulation or deleting it
	($4 \to 1$), respectively. Note that for these operations to be 
	possible, the point E must lie within the simplex (A, B, C, D).}
\end{figure}

From the Radon partition in ${\mathbb R}^3$ one finds that there exist
four possible flips in three dimensions, two for every partition in figure 
\ref{Fradon}. 
For the case of figure \ref{Fradon}a the two possible flips are shown in figure 
\ref{Fflip14}.
The flip changing the triangulation from $1$ to $4$ simplices corresponds to
adding a vertex to an existing triangulation. Note however, that in practice
the inverse transformation may not always be applicable, since the 
configuration of one vertex ($E$ in figure \ref{Fflip14}) being the endpoint of
exactly four simplices is rarely ever present in a triangulation.
This fact -- in combination with the existence of non-tetrahedralizable 
polyhedra in three dimensions -- complicates the deletion of vertices
from triangulations \cite{shewchuk2002}.

\begin{figure}[ht]
  \centering {\includegraphics[height=8cm]{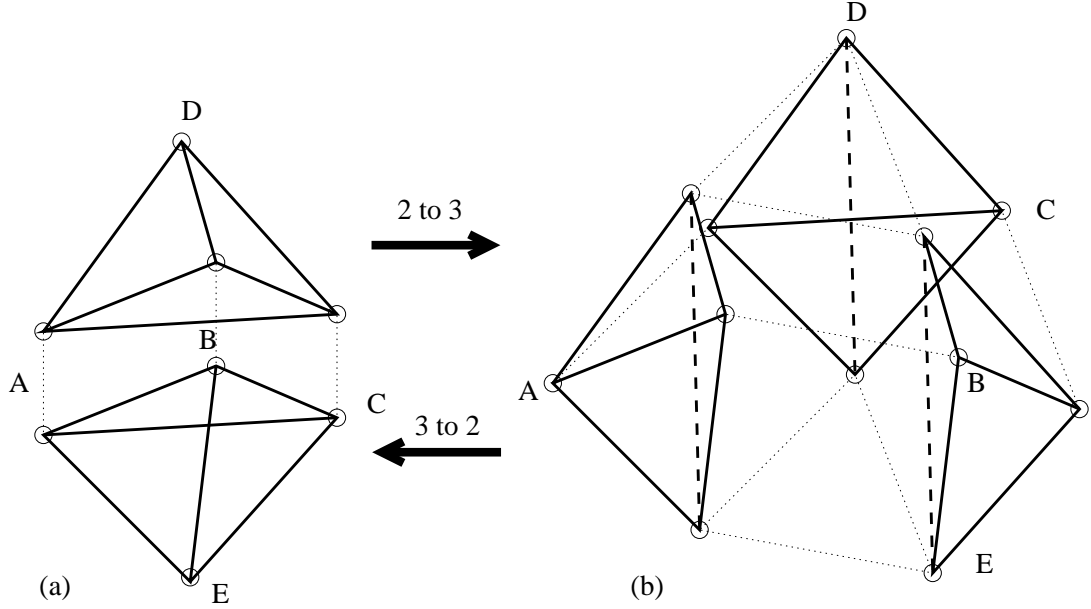}}
  \caption{\label{Fflip23}
	Three-dimensional illustration of the possible triangulations of 
	five points. In (a) there are two simplices:
	(A, B, C, D), (A, B, C, E), sharing the common triangle (A, B, C), 
	whereas the right picture (b) consists of three simplices 
	(A, B, D, E), (B, C, D, E), (C, A, D, E). The simplices have been
	taken apart for clarity and the dotted lines have been drawn to 
	connect the identical points. Invisible Edges have been drawn with dashed lines. 
	Note that the flips can only be performed, if the polyhedron 
	(A, B, C, D, E) is convex, since otherwise the flips will result in
	overlaps with additional neighboring simplices (not shown here).}
\end{figure}

The second partition in figure \ref{Fradon}b following from Radon's theorem requires a more careful
evaluation, see figure \ref{Fflip23}.
The flips $2 \to 3$ (and $3 \to 2$) can only be performed
if the polyhedron formed by the five points in ${\mathbb R}^3$ is convex, 
otherwise the operation would yield overlapping simplices in the triangulation.
The convexity of $A, B, C, D, E$ in figure \ref{Fflip23} can be tested by 
checking if for every edge $A,B$ and $B,C$ and $C,A$ there exists a 
hyperplane which has the remaining three points ($D, E, A/B/C$) on the same 
side \cite{Edelsbrunner1996,muecke1998,Sauer1995}.

\subsection{The Delaunay Criterion}\label{SSdelaunay}

Every tetrahedron in ${\mathbb R}^3$ has a uniquely defined circumsphere, if 
the four vertices do not lie on a common plane (i.e.~if the tetrahedron is not 
flat). 
The Delaunay triangulation is a triangulation where all the simplices satisfy
the {\it Empty-Circumsphere-Criterion}:
No vertex of the triangulation may lie inside the circumspheres of the 
triangulation simplices.
Thus, the Delaunay triangulation is uniquely defined if the vertices are in 
general position, i.e.~if no five vertices must lie on a common sphere and no
four vertices may lie on a common plane \cite{Fortune1992}.

The simplest method to determine, whether a vertex $V$ lies outside or inside the
circumsphere of a simplex $(A, B, C, D)$ is to solve the associated four 
sphere equations. However, this problem can be solved more efficiently by 
adding one more dimension \cite{Ferrez2001,Goodman1997,muecke1998}.

Suppose we would like to know whether the vertex $E$ lies in- or outside the
circumsphere of the simplex $(A, B, C, D)$, which we will -- without loss of
generality -- assume to be positively oriented. 
Then one can proceed as follows (e.g. \cite{aurenhammer1988}):
Project the coordinates in ${\mathbb R}^3$ onto a paraboloid in 
${\mathbb R}^4$ via
\bea
  A=(A_x, A_y, A_z) \to A^+ = (A_x, A_y, A_z, A_x^2+A_y^2+A_z^2)\,.
\eea
The four points $A^+, B^+, C^+, D^+$ define a hyperplane in ${\mathbb R}^4$.
If $E$ is within the circumsphere of $(A, B, C, D)$, then $E^+$ will be below
this hyperplane in ${\mathbb R}^4$ and above otherwise.
Consequently, the in-circumsphere-criterion in ${\mathbb R}^3$ reduces to a 
simple orientation computation in ${\mathbb R}^4$, i.e.~by virtue of this
lifting transformation one finds \cite{Ferrez2001}
\bea\label{Einsphere}
  && {\rm in\_circumsphere}((A, B, C, D), E) \nn\\ 
  & = &	{\rm oriented}(A^+, B^+, C^+, D^+, E^+) \nn\\
  & = & {\rm sign} \left| \begin{tabular}{ccccc}
	$A_x$ & $A_y$ & $A_z$ & $A_x^2+A_y^2+A_z^2$ & $1$\\
	$B_x$ & $B_y$ & $B_z$ & $B_x^2+B_y^2+B_z^2$ & $1$\\
	$C_x$ & $C_y$ & $C_z$ & $C_x^2+C_y^2+C_z^2$ & $1$\\
	$D_x$ & $D_y$ & $D_z$ & $D_x^2+D_y^2+D_z^2$ & $1$\\
	$E_x$ & $E_y$ & $E_z$ & $E_x^2+E_y^2+E_z^2$ & $1$ 
	\end{tabular}\right|\\
  & = & {\rm sign} \left| \begin{tabular}{cccc}
	$A_x-E_x$ & $A_y-E_y$ & $A_z-E_z$ &
	$(A_x^2+A_y^2+A_z^2)-(E_x^2+E_y^2+E_z^2)$\\
	$B_x-E_x$ & $B_y-E_y$ & $B_z-E_z$ & 
	$(B_x^2+B_y^2+B_z^2)-(E_x^2+E_y^2+E_z^2)$\\
	$C_x-E_x$ & $C_y-E_y$ & $C_z-E_z$ & 
	$(C_x^2+C_y^2+C_z^2)-(E_x^2+E_y^2+E_z^2)$\\
	$D_x-E_x$ & $D_y-E_y$ & $D_z-E_z$ & 
	$(D_x^2+D_y^2+D_z^2)-(E_x^2+E_y^2+E_z^2)$
	\end{tabular}\right|\nn\,,	
\eea
where a positive sign is to be taken as an affirmative 
answer\footnote{In the general case one will have to multiply by the orientation of $(A, B, C, D)$.}.

Several algorithms have been developed for the construction of static
triangulations \cite{Goodman1997,devillers1998} as well as for the maintenance of 
kinetic triangulations \cite{Ferrez2001,Xinjian1997}, some
of which will be discussed in section \ref{Simplementation}.

\subsection{The Geometric Dual: Voronoi Tessellation}\label{SSvoronoi}

The most general Voronoi tessellation (sometimes also called Dirichlet tessellation) of a 
set of generators $\{c_i\}$ in ${\mathbb R}^d$ is defined as a partition of 
space into regions $V_i$:
\bea\label{Evoronoi}
  V_i & = & \{x \in \mathbb{R}^n \; : \; \mathcal{P}(x,c_i) \leq \mathcal{P}(x,c_j)
	\qquad \forall j \neq i \}\,,
\eea
where $\mathcal{P}(x,c_i)$ can be an arbitrary function, which reduces in the standard case 
of the simple Voronoi tessellation to the normal euclidian distance 
${\mathcal P}(x,c_i)=\abs{x-c_i}$. In other words, the 
Voronoi cell around the generator $c_i$ contains all points in ${\mathbb R}^d$
that are closer to $c_i$ than to any other generator $c_j$. Note that this 
partition is -- unlike the Delaunay triangulation -- uniquely defined also
for point sets that do not fulfill the general position assumption.
Voronoi tessellations have many interesting applications in practice -- for
a survey see e.g.~\cite{okabe2000} -- since they do describe
influence regions.

\begin{figure}[ht]
  \centering {\includegraphics[height=6cm]{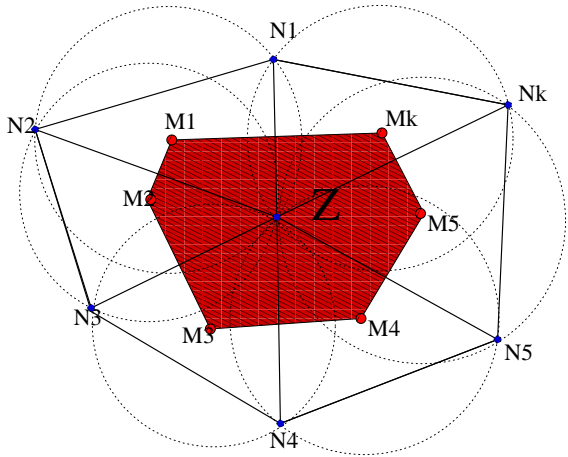}}
  \caption{\label{Fvoronoi2d}
	Two-dimensional Voronoi cell around a generator $Z$, which is 
	surrounded by other generators $N_i$. All points within the shaded 
	region are closer to the generator Z than to any other generator. 
	The corners of the Voronoi cell polygons $M_i$ are the centers of the 
	circumcircles (drawn with dotted lines) of the Delaunay triangulation
	(solid lines) of the generator set.}
\end{figure}

In 2 dimensions Voronoi cells are convex polygons completely 
covering the plane, see e.g.~figure \ref{Fvoronoi2d}.
This finding generalizes to arbitrary dimensions: The boundaries between two
$d$-dimensional Voronoi regions $V_i$ and $V_j$ as defined in 
(\ref{Evoronoi}) reduce to the equation for a $d-1$ hyperplane.
Therefore per definition the Voronoi cells around generators $Z_i$ 
situated on the convex hull of the point set $Z=\{Z_1, Z_2, \ldots, Z_n\}$ 
will extend to infinity and thus will have an infinite volume. 

Voronoi tessellations can be constructed like the well-known
Wigner-Seitz cell in solid state physics \cite{Kittel1996}, but fortunately
there are much more efficient ways to construct the Voronoi 
tessellation.
In this work, we will exploit the geometric duality with the Delaunay triangulation:
In any dimension, the corners of the Voronoi polyhedra are the centers of the 
circumspheres of the $n$-simplices contained in the Delaunay triangulation of 
the Voronoi generators, see figure \ref{Fvoronoi2d}.

The introduction of influence regions also enables the definition of proximity 
between vertices:
We understand two vertices to be direct neighbors (in the sense that their 
influence regions touch) if they share a common face in their Voronoi diagram 
or -- equivalently -- if they are direct neighbors in the dual Delaunay 
triangulation, see figure \ref{Fvoronoi2d}. 

The concept of the Voronoi cell can easily be extended towards generators with a 
varying strength, the weighted Voronoi tessellation \cite{okabe2000}.
In such extensions, every generator is assigned a weight, i.e.~the 
functions $\mathcal{P}(x,c_i)$ in (\ref{Evoronoi}) are then given by a function 
describing the influence strength of the generator $i$ at $c_i$ on $x$. 
Obviously, the weighted Dirichlet regions can -- in contrast to the standard unweighted 
case -- be empty, e.g.~if a vertex with a weak influence is surrounded by 
strong vertices.
Among many possible choices for weight functions \cite{okabe2000,Aurenhammer1991} 
we will explicitly mention here the case of power-weighted Voronoi diagrams, 
also often called the Laguerre complex \cite{Ferrez2001,Xinjian1997}. 
It is obtained by assigning a weight 
$\omega_i \in \mathbb{R}$ to every generator $c_i$, i.e.~by using
$\mathcal{P}(x,c_i) \to \mathcal{P}(x,c_i,\omega_i)=(c_i-x)^2-\omega_i^2$ in
(\ref{Evoronoi}).
In the power-weighted case one can still show that in three dimensions the 
Laguerre cells are convex polyhedra, whose corners can be obtained from the 
weighted centers of the corresponding weighted Delaunay triangulation 
tetrahedra -- where the empty circumsphere criterion is simply replaced by
its weighted counterpart. Therefore,
the Laguerre tessellation or its geometric dual -- the weighted Delaunay triangulation -- 
is suitable for collision detection between differently sized spheres.
We would like to stress that the algorithms in this article can be generalized to 
the power-weighted case in a straightforward way.

\subsection{Voronoi surfaces and volumes}\label{SSsurface_volume}

\begin{figure}[ht]
  \centering {\includegraphics[height=6cm]{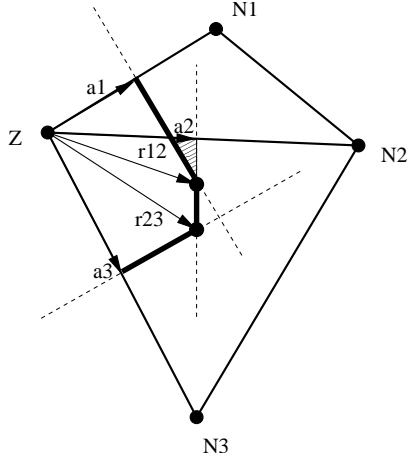}}
  \caption{\label{Fvoronoi_volume}
	Volume (area) computation of a two-dimensional Voronoi cell around the
	generator $Z$.
	Here for clarity only two simplices are shown. The first simplex
	$(Z, N_1, N_2)$ contributes the positive area spanned by 
	$(a_1, r_{12})$ and the negative area spanned by $(r_{12}, a_2)$. The
	second contribution is negative because the center of the circumcircle
	of $(Z, N_1, N_2)$ is outside the simplex. Most of the negative
	volume contribution is thus canceled -- up to a small part 
	(hatched region) situated outside the Voronoi cell boundary 
	(thick lines). However, when considering
	the next simplex $(Z, N_2, N_3)$ the positive contribution spanned by
	$(a_2, r_{23})$ cancels with this remaining negative contribution from
	the first simplex. The last contribution comes from $(r_{23}, a_3)$ 
	and is again positive.}
\end{figure}

Within the framework of growth models \cite{Xinjian1997}, tissue simulations
\cite{Weliky1990,Weliky1991} and the solution of partial differential 
equations on irregular grids \cite{Miller1995,Bottino2000}, not only the 
neighborship relations in the Delaunay triangulation but also the 
corresponding Voronoi cell volumes as well as the contact surface between two 
Voronoi cells may become important.
Obviously, to the surface of a Voronoi cell around a vertex $Z$ every incident simplex $\sigma$ with 
$Z \in \sigma$ contributes, see figure \ref{Fvoronoi2d}. In two dimensions
every triangle $(Z, N_i, N_{i+1})$ contributes two surfaces, spanned by the 
half distances between the vertices $A=1/2(N_i-Z)$ and $B=1/2(N_{i+1}-Z)$, and
the center of the circumcircle of $R=COC(Z, N_i, N_{i+1})-Z$.
If $(A,B)$ are positively oriented (which we will further on assume without loss
of generality), then the oriented 2-volume contribution of the simplex 
$(Z, N_i, N_{i+1})$ to the Voronoi cell volume of the vertex $Z$ is given by
\bea\label{Evolume2d}
  V_Z=\frac{1}{2}\left(
	\left|\begin{tabular}{cc}
		$A_x$ & $R_x$\\
		$A_y$ & $R_y$ \end{tabular} \right|
	+\left|\begin{tabular}{cc}
		$R_x$ & $B_x$\\
		$R_y$ & $B_y$\end{tabular}\right|\right)\,,
\eea
see also figure \ref{Fvoronoi_volume} for illustration. Obviously, the sum of
the three volumes has to equal the total simplex volume
$V_Z+V_{N_i}+V_{N_{i+1}}=V_S$ . One can show algebraically, that this identity
holds true for any $R$, i.e.~also in the interesting case where $R$ being the
center of the circumcircle lies outside the triangle as in 
figure \ref{Fvoronoi_volume}.
In this case one of the two area contributions in (\ref{Evolume2d}) will be
negative.
Now when considering figure \ref{Fvoronoi_volume} it becomes clear that by
adding the oriented volume contributions of all simplices containing the 
vertex $Z$ one obtains the correct volume for the Voronoi cell around $Z$.

This finding generalizes to three-dimensional volume computation, and the
three-dimensional contact surface contribution is in fact nothing but a 
two-dimensional volume computation, since the contact surfaces between two Voronoi 
polyhedra are plane polygons.

The algorithm leads to a wrong Voronoi cell volume at the 
boundary, where the Voronoi cells are per definition infinite.
However, if the centers of the circumspheres of the simplices at the boundary
do not lie outside the convex hull of the triangulation, then the volume
summation yields the part of the Voronoi cells which is inside the convex
hull. This is desirable for some configurations, such as the solution of
partial differential equations \cite{Bottino2000}.

The numerical complexity of the volume computation is linear with 
the number of simplices surrounding the vertex, whereas the complexity of
contact surface calculation between two generators grows linear with the 
number of simplices containing both generators as endpoints.
The algorithm can be checked by using that the sum over all Voronoi cell volumes computed this way
has to equal the sum of the simplex volumes in the dual Delaunay triangulation.

\section{Algorithms and Implementation}\label{Simplementation}

\subsection{The Data Structure}\label{SSdatastructure}

As has already been mentioned in the introduction, three-dimensional Delaunay
triangulations are much more complicated than in two dimensions. The first 
difference is that the number of triangulation simplices may vary for a 
constant number of mobile vertices. 
Our triangulation basically consists of two data structures:
\begin{itemize}
\item{a list of the vertices in the triangulation}
\item{a list of connected 3-simplices (tetrahedra) contained in the triangulation}
\end{itemize}
Both structures are organized in a list to enable for the kinetic movement,
insertion, and deletion of vertices and the corresponding dynamic update of the 
simplex list.

A vertex consists of the $x$, $y$ and $z$ coordinates, a weight (which will be 
assumed to be equal for all points throughout this article) and -- 
to compute the Voronoi cell data efficiently -- a vector of all incident
tetrahedra.
A simplex consists of four pointers on vertices and of four pointers on the
neighboring simplices. The latter is required by the fact that we perform a walk in
the triangulation (see section \ref{SSlocation}). 

The construction of the Delaunay triangulation basically relies on two basic
predicates: The determination whether two points lie on the same side of a 
plane defined by three others and the question whether a point lies in- or 
outside the circumsphere circumscribing the simplex of four others 
(enforcement of the Delaunay criterion).
By using these two simple predicates the whole triangulation can be built up.

\subsection{Incremental Insertion Algorithms}\label{SSincinsert}

Unfortunately in Delaunay triangulations the insertion of one
new vertex can change the whole triangulation, but this only holds true for 
some extreme vertex configurations, for some examples see 
\cite{okabe2000}. In practice, the effect of adding a new vertex to a
Delaunay triangulation will nearly always be local. Anyhow, the algorithms we
describe here can of course also cope with these worst case scenarios.

\begin{figure}[ht]
  \centering {\includegraphics[height=6cm]{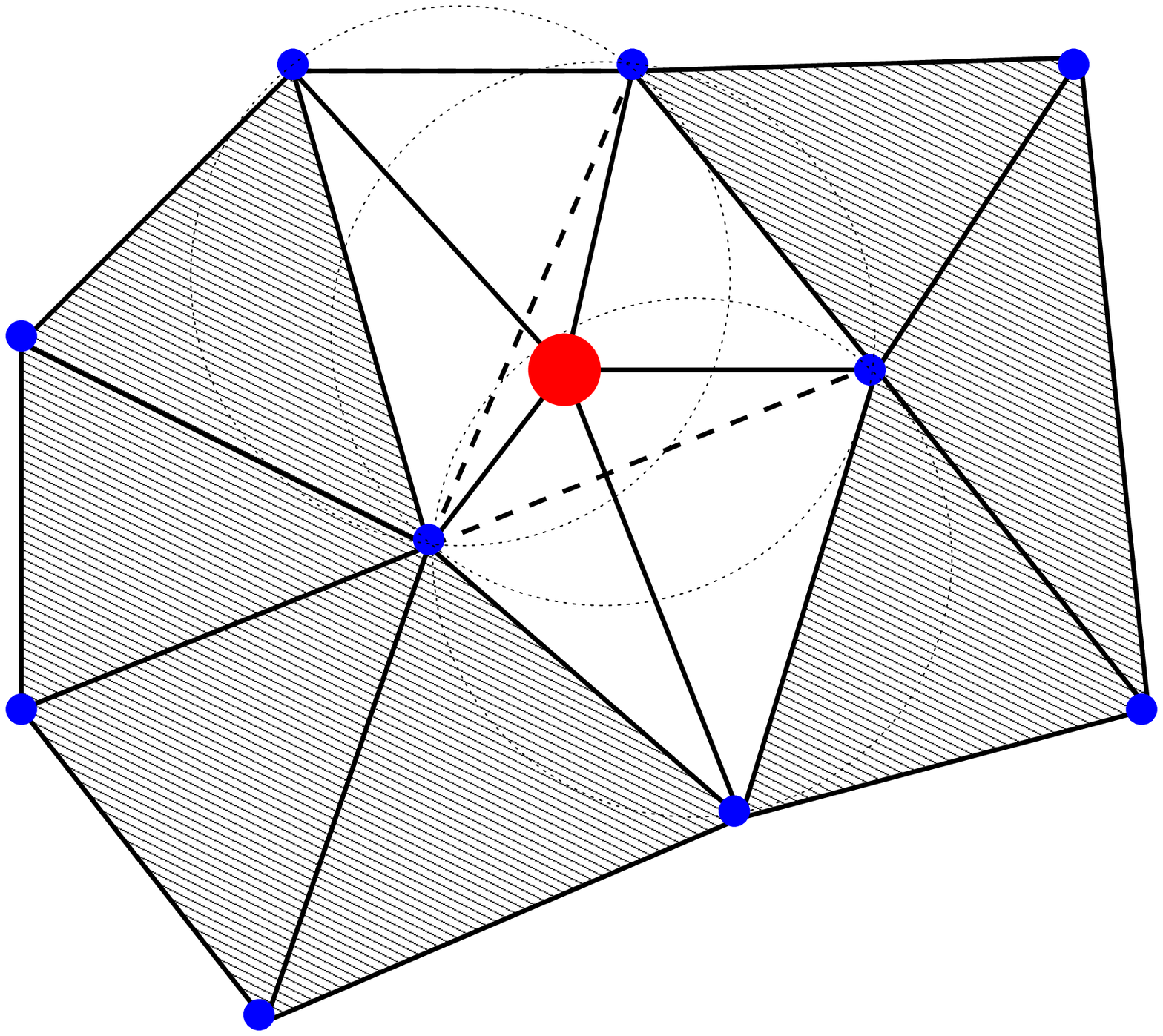}}
  \caption{\label{Finsertion}
	In this example, a new vertex (large point) is inserted into an 
	existing triangulation (not all simplices are shown). Most of the 
	simplices remain valid (shaded region), but 3 simplices (dashed lines)
	contain the new vertex within their circumspheres (dotted lines). These
	are replaced by 5 new simplices (solid lines) formed by the new vertex
	and the external facets (lines in two dimensions). 
	The resulting triangulation fulfills the Delaunay criterion. }
\end{figure}

So let us assume we have a valid Delaunay triangulation with $n$ vertices. Let
us furthermore assume that the new vertex lies within the convex hull of the
$n$ vertices. Then the updated Delaunay triangulation can be constructed as
follows (see figure \ref{Finsertion}):
\begin{itemize}
\item{Identify all invalid simplices in the triangulation, i.e.~all those 
	containing the new vertex within their circumsphere.}
\item{Collect the external facets of the invalid simplices. (Those are the
	triangles facing valid simplices.)}
\item{Replace the invalid simplices by new ones formed via combining the
	external facets with the new vertex.}
\end{itemize}
Sometimes this incremental algorithm is also called 
{\it Bowyer-Watson Algorithm}
\cite{Boissonnat1993,Choi1997}. Once all the invalid simplices have been found,
its computational cost is very low (linear with the total number of 
invalid simplices). At first it actually suffices to find the one simplex which
contains the new vertex within its convex hull -- the remaining simplices can be 
found by iteratively checking all neighbors for invalidity, see also figure \ref{Finsertion}.
The result of this procedure is a Delaunay triangulation with $n+1$ vertices.

The algorithm shown in figure \ref{Finsertion} is only slightly different than the 
{\it Green-Sibson Algorithm} \cite{Boissonnat1993}, which needs the simplex 
containing the new vertex as an input. Then the elementary topological transformation $1 \to 4$ is 
performed with this simplex and the resulting (Delaunay-invalid) triangulation
is transformed to a Delaunay triangulation by performing $2 \to 3$ and 
$3 \to 2$ flips until all simplices fulfill the Delaunay property.

Note that for weighted triangulations the simplex containing the new 
vertex within its convex hull is not necessarily invalid, since the weighted
circumsphere does not generally contain the complete simplex. This corresponds to
the case of an empty Laguerre cell -- the vertex therefore has to be rejected.

The initial triangulation can be given by an artificial large simplex which contains all 
the data to be triangulated within its convex hull. Therefore, the convex hull of the points
to be triangulated is contained within the artificial simplex and is not reproduced by the 
triangulation. However, in the framework of kinetic proximity structures this has the advantage
that one does not have the problem of maintaining the convex hull of moving points, since the
artificial simplex does not move.
The initial simplex must therefore be large enough to contain the data within its insphere
throughout the full time evolution of the simulation.
One choice for such an initial simplex is a $\rm{CH}_4$ configuration. 

\subsection{Location of Simplices}\label{SSlocation}

The incremental insertion algorithm requires a first initial simplex containing the
new vertex. Many all implementations of Delaunay triangulations perform a walk in 
the triangulation, for an overview of different walking strategies see e.g. \cite{devillers2001}.
Note that points can also be located by using the history the triangulation has been constructed
(e.g.~the so-called Delaunay tree \cite{Boissonnat1993} or history dag \cite{Facello1993}). However, we
are aiming at kinetic triangulations, where the length of a history stack could not be controlled.

\begin{figure}[ht]
  \centering {\includegraphics[height=8cm]{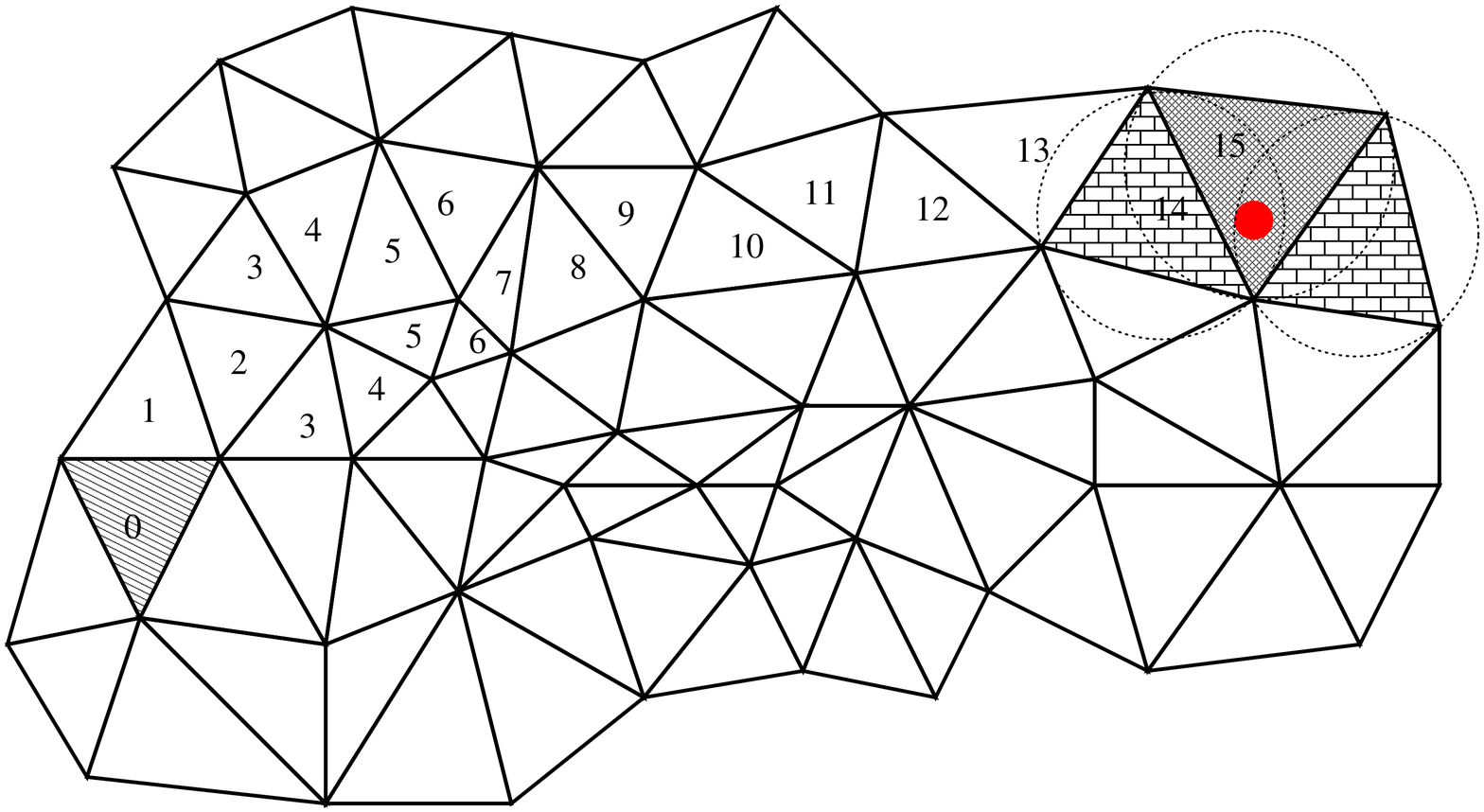}}
  \caption{\label{Fhopping}
	Two-dimensional illustration of the hopping algorithm in a 
	triangulation (not all simplices are shown). Starting from
	the hatched initial simplex $0$ the algorithm finds a way towards the
	invalid simplex $15$ (cross-hatched) that contains the new vertex 
	(large point). 
	The computational time necessary for the walk algorithm is proportional
	to the number of traversed simplices.}
\end{figure}

Here we will use a stochastic visibility walk \cite{devillers2001} to locate a simplex containing 
a point. 
Starting with an arbitrary initial simplex $A$ and a new vertex $v$ to be 
inserted in the triangulation, in the normal visibility walk one of the four neighbor simplices of
$A$ is chosen using the following criterion:

\begin{itemize}
\item{For all four vertices $a_{i=1,2,3,4}$ of the simplex $A$ check with the 
	new vertex $v$:\\
	Are the vertices $a_i$ and $v$ on different sides of the plane defined
	by the other three vertices $a_{j \neq i}$, i.e. is it visible from the
	plane defined by the $a_{j \neq i}$?\\
	yes : $\Longrightarrow$ Jump to the simplex opposite to $a_i$.}
\item{If no neighbor simplex is found, the vertex $v$ is contained within
	the simplex $A$ and the destination is thus reached.}
\end{itemize}
Obviously, the algorithm can take different pathways 
(see figure \ref{Fhopping}) since there may be more
than one neighbor fulfilling this criterion. 
Note also that due to numerical errors the normal visibility walk may loop when
triangulating regular lattices (such as cubic, ...) that violate the general
position assumption. Such situations can be easily avoided by using the stochastic 
visibility walk, where the order of the vertices to be checked is randomized. Such a stochastic
visibility walk terminates with unit probability \cite{devillers2001}.

The complexity is directly proportional to the length of the path to be walked -- measured
in units of traversed simplices. For $n$ uniformly distributed vertices for 
example, the average total number of simplices will grow linearly ($n$) with the number 
of vertices, whereas the average distance between two arbitrarily selected 
simplices will grow like $n^{1/3}$. Once the invalid simplex has been found,
the average remaining complexity for the incremental insertion will be in average constant (in $n$).
Therefore one would expect the overall theoretical complexity to behave like 
$\alpha n^{4/3}+\beta n$ for uniformly distributed points and in higher dimensions $d$ as
$\alpha n^{1+1/d}+\beta n$, see also \cite{Bowyer1981}.

Obviously, the algorithm heavily depends on a good choice of the starting 
simplex. The method could therefore be improved by checking whether the new 
vertices lies within a certain subregion which means preprocessing, or it can be
speeded up by initially using larger step sizes, e.g. by using several triangulations of 
subsets of vertices \cite{devillers1998}. Alternatively, one can choose the closest vertex
out of a random subset of the triangulation to find a good starting simplex \cite{muecke1996,devroye1998}.
In some sense these two methods are similar, since finding a close vertex is nothing but finding 
the Voronoi-cell the point is situated in. The second method however does not require
the maintenance of an additional triangulation in the case of kinetic vertices.

In many practical simulations, some neighborship relations may already be 
known when building the initial triangulation. Our implementation 
of the incremental algorithm therefore expects the vertices to be included in some
order, such that successive vertices are also very close to each other in the
final triangulation and therefore chooses the starting simplex in the walk 
algorithm as being the last simplex created if no other guess is given.
Note that for processes as cell proliferation, the situation is even better: 
New cells can be created
by cell division which corresponds to the insertion of a new vertex close to
an existing one. Consequently, one always has a perfect guess for the starting 
simplex in these cases.

\subsection{Updating the triangulation}\label{SSflips}

In three-dimensional kinetic triangulations where the vertices are moving and neighborship
relations can change, one has to deal with a changing number of simplices.
Once deletion and insertion of points has been implemented, a simple method handling 
kinetic vertices would be to delete them at their old position and perform an insertion at
the new position \cite{fabritiis2003,lauritsen1994}. However, since these operations involve
many simplices, there exist more efficient algorithms.

It is evident that in the case of moving vertices the Delaunay criterion may
be violated, i.e.~after the vertices have moved one may end up with a 
triangulation that violates the Delaunay criterion. Even worse, if the 
vertices move too far, e.g.~if one vertex moves inside another simplex, the 
triangulation will become invalid (contain overlapping simplices). This must 
be avoided by either computing a maximum step size (see e.g.~section 
\ref{SSdeletion}) or by simply keeping the displacements safely small.
So let us assume here that after vertex movement one is left with a valid
triangulation which possibly violates the Delaunay criterion.
Recomputing the whole triangulation is usually not an option for large data sets.
The elementary topological transformations in subsection \ref{SSett} however
can be exploited to restore the Delaunay criterion. Since we will neither add 
nor delete vertices in this subsection it is evident that the flips 
$1 \to 4$ and $4 \to 1$ are not necessary\footnote{This will change for weighted triangulations,
as vertex movement may lead to trivial vertices that have no associated Laguerre cell volume and
vanish from the triangulation.}. Consequently, the transformations $2 \to 3$ and $3 \to 2$
will suffice to transform the given triangulation into a Delaunay 
triangulation, which has been shown to work \cite{Ferrez2001,Edelsbrunner1996,muecke1998}.
With a glance at figure \ref{Fflip23} one can see that indeed the flip $2 \to 3$ 
effectively creates a neighborship connection, whereas the flip $3 \to 2$
destroys it.
Therefore we have implemented routines to check either the complete list of simplices
or only a small subset for Delaunay invalidity.
Our simple data structure enables a convenient calculation of the flip 
criteria in three dimensions in subsection \ref{SSett}.
The main advantage of the flip algorithm is that it is -- in average -- 
linear in the number of simplices which is also linear with the number of vertices
in most practical applications.

We iterate through the list of simplices and check for flipping-possibilities among every
simplex (the active simplex) and its neighbors (the passive simplices):
Given a simplex $S$ and its neighbor $N_i$, the flip $2 \to 3$ is 
performed if the following two conditions are met:
\begin{itemize}
\item{The opposing vertex of the neighbor $N_i$ lies within the circumsphere
	of $S$.}
\item{The five points in the union of the two simplices form a convex polyhedron.}
\end{itemize}
Due to Radon's theorem it suffices in the last criterion to check whether
the edges of the common triangle (without loss of generality $(S_A, S_B, S_C)$)
are convex with respect to the other two points ($S_D$ and $N^i_{opp}$)
\cite{Edelsbrunner1996,muecke1998}.

The criterion for the flip $3 \to 2$ can be written as follows:
Given the simplex $S$ and two of its neighbors $N_i$ and $N_j$, the flip
$3 \to 2$ is performed if the following conditions are met:
\begin{itemize}
\item{The simplex $N_i$ is also a neighbor of $N_j$.}
\item{The neighbor pairs $(S, N_i)$,$(S, N_j)$ and $(N_i, N_j)$ 
	violate the Delaunay criterion.}
\end{itemize}
If any flip is performed, the new simplices must be inserted at the end of the
list of simplices to be checked again. 
The algorithm terminates as the end of this list is reached.

Note however, that for these flips to be possible, all simplices must be 
disjoint (the intersection of two simplices may at most be a triangle), i.e.
the triangulation must be valid -- flips cannot be used to recover from an invalid triangulation.
This becomes an issue when computing a maximum step size for the vertex kinetics,
compare subsection \ref{SSdeletion}.
  
\subsection{Deletion of Vertices}\label{SSdeletion}

In many problems (e.g.~mesh generation) the deletion of vertices from a
Delaunay triangulation is not of great importance, since there is no great 
advantage other than a negligible gain in efficiency. However, if the 
triangulation is used for example for proximity structures or data interpolation, 
vertex deletion may become important.

Several algorithms have been developed to manage the deletion of vertices in
two dimensions, see e.g.~\cite{Brouns2001,Devillers2002,mostafavi2003}. 
As has already been argued, there exist some fundamental differences between 
the two-dimensional and the higher-dimensional case. 
Simply removing a vertex together with its incident simplices leaves a 
star-shaped hole in the triangulation, which is not necessarily convex.
Unlike in two-dimensional case, where a star-shaped polygon always admits a
triangulation which can be transformed by flips into the Delaunay triangulation
\cite{Brouns2001,Devillers2002} in three dimensions a 
star-shaped polyhedron may not admit a tetrahedralization. The simplest 
example for such a polyhedron is Sch\"onhardt's polyhedron 
\cite{schoenhardt1928}, reported among others in \cite{shewchuk2002,Boissonnat1998}.
The star-shaped holes emerging in Delaunay triangulations however, will always possess a
tetrahedralization, which has been proven in \cite{shewchuk1998}. Note however, that this
does not generally hold true for constrained Delaunay triangulations \cite{shewchuk2002}.

Another approach for deletion is given in \cite{Vigo2000}, where the history is used to 
reconstruct the triangulation such that the vertex has never been inserted. 
Again, in our approach we did not want to use some kind of history, since for kinetic
triangulations there is no way to control the size of the history.

\begin{figure}[ht]
  \centering {\includegraphics[height=4cm]{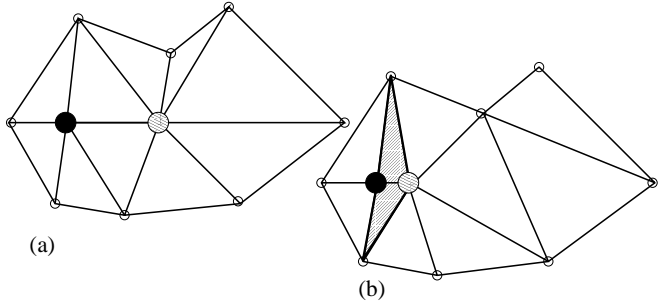}}
  \caption{\label{Fdeletion}
	Two-dimensional illustration of vertex deletion from a Delaunay 
	triangulation. In part (a), the vertex to be deleted (large hatched 
	point) is moved	in several steps followed by flips restoring the
	Delaunay property towards its closest neighbor 	(large solid point), 
	until the inner simplices (shaded region) can be safely deleted 
	[part (b)]. The two vertices are simply merged and the remaining opposing
	simplices are connected as neighbors. Finally, the Delaunay criterion 
	is again restored by using flips.}
\end{figure}

The basic idea of our approach to delete a vertex is to move it towards
its nearest neighbor in several steps, each followed by a sequence of flips
$2 \to 3$ and $3 \to 2$ restoring the Delaunay property,
until the simplices between the two vertices are very flat and can be clipped
out of the triangulation without harming its validity. In some sense we
project the problem of vertex deletion on the already presented algorithm for
vertex movement. Figure \ref{Fdeletion} illustrates the idea of the algorithm.

The main questions to be answered all reduce to the problem of the step size. 
How far can a vertex $v_i$ be moved into a certain direction without invalidating
the triangulation, i.e.~without creating overlapping simplices?
If the vertex $v_i$ penetrates another simplex, the orientation of at least
one of its surrounding simplices will change. Therefore one can derive a 
step size criterion by demanding that the orientation of the simplices 
incident to $v_i$ may not change sign. We define the pseudo-orientation of a 
simplex $S_i=(A^{(i)}, B^{(i)}, C^{(i)}, D^{(i)})$ as follows:
\bea
  {\mathcal V}_0^{(i)} & = & \left| \begin{tabular}{cccc}
	$A_x^{(i)}$ & $A_y^{(i)}$ & $A_z^{(i)}$ & $1$\\
	$B_x^{(i)}$ & $B_y^{(i)}$ & $B_z^{(i)}$ & $1$\\
	$C_x^{(i)}$ & $C_y^{(i)}$ & $C_z^{(i)}$ & $1$\\
	$D_x^{(i)}$ & $D_y^{(i)}$ & $D_z^{(i)}$ & $1$
	\end{tabular}\right|\nn\\
	& = & \left| \begin{tabular}{ccc}
	$A_x^{(i)}-B_x^{(i)}$ & $B_x^{(i)}-C_x^{(i)}$ & $B_x^{(i)}-D_x^{(i)}$\\
	$A_y^{(i)}-B_y^{(i)}$ & $B_y^{(i)}-C_y^{(i)}$ & $B_y^{(i)}-D_y^{(i)}$\\
	$A_z^{(i)}-B_z^{(i)}$ & $B_z^{(i)}-C_z^{(i)}$ & $B_z^{(i)}-D_z^{(i)}$
	\end{tabular}\right|\,.
\eea
In the second line we have reordered the terms such that the vertex to be
moved is in the first column.
In fact, this is -- up to a factor of $1/6$ -- the signed volume of the 
simplex $S_i$. Now suppose that one of the vertices -- without loss of 
generality we have chosen $A$ -- is moved along 
the direction of $\Delta$, i.e.~$A \to A^\prime = A + \lambda_i \Delta$ with
$\lambda \in {\mathbb R}$ and $\Delta=(\Delta_x, \Delta_y, \Delta_z)$. Then 
the new pseudo-orientation is obtained via
\bea
  {\mathcal V}_1^{(i)} = {\mathcal V}_0^{(i)} + \lambda_i
	\left|\begin{tabular}{ccc}
	$\Delta_x$ & $B_x^{(i)}-C_x^{(i)}$ & $B_x^{(i)}-D_x^{(i)}$\\
	$\Delta_y$ & $B_y^{(i)}-C_y^{(i)}$ & $B_y^{(i)}-D_y^{(i)}$\\
	$\Delta_z$ & $B_z^{(i)}-C_z^{(i)}$ & $B_z^{(i)}-D_z^{(i)}$
	\end{tabular}\right|\,.
\eea
If the orientation of the simplex $S_i=(A_i, B_i, C_i, D_i)$ is not allowed 
to change one has found an upper bound for $\lambda_i$ via
\bea\label{Econdition}
  \lambda_i(S_i) = \frac{\abs{\left({\mathcal V}_0^{(i)}\right)}}
	{{\rm abs}\left|\begin{tabular}{ccc}
	$\Delta_x$ & $B_x^{(i)}-C_x^{(i)}$ & $B_x^{(i)}-D_x^{(i)}$\\
	$\Delta_y$ & $B_y^{(i)}-C_y^{(i)}$ & $B_y^{(i)}-D_y^{(i)}$\\
	$\Delta_z$ & $B_z^{(i)}-C_z^{(i)}$ & $B_z^{(i)}-D_z^{(i)}$
	\end{tabular}\right|}\,.
\eea
Of course this check has to be done for all simplices incident to the 
moving vertex $A$, i.e.~with
\bea\label{Eoverall}
  \lambda = \min_{S_k : A \in S_k} \lambda_k
\eea
one has an overall measure of the maximum step size of $A$ in the direction
of $\Delta$. If $\lambda > 1$, then the vertex can simply be moved along the 
complete path $(\Delta_x, \Delta_y, \Delta_z)$, whereas if $\lambda < 1$ the 
vertex $A$ can only be moved by a fraction $\alpha \Delta \;:\;\alpha<\lambda$. 
Let us furthermore define $A^\prime$ to be the nearest neighbor of $A$. These
vertices will have a certain number of simplices in common. 
For the remaining simplices we define the quantity $\lambda_{\rm REST}$ 
in analogy to $\lambda$ via
\bea\label{Erest}
  \lambda_{\rm REST}=\min_{S_k : A \in S_k, A^\prime \notin S_k} \lambda_k
\eea
Thus, our algorithm for deleting a vertex $A$ can be summarized as follows:
\begin{enumerate}
\item{Find the nearest neighbor vertex $A^\prime$.}
\item{{\it repeat}
	\begin{itemize}
	\item{set $\Delta=A^\prime-A$}
	\item{determine $\lambda=\min_{S_k \; : \; A \in S_k} \lambda_k$}
	\item{determine 
	$\lambda_{\rm REST}=\min_{S_k \; : \; A \in S_k, A^\prime \notin S_k} 
	\lambda_k$}
	\item{if $\lambda_{\rm REST} \le 1.0$ move 
	$A \to A+\alpha \lambda \Delta$ with 
	$\alpha<1$ and update the simplices surrounding $A$ with 
	flips to restore Delaunay property}
	\end{itemize}
	{\it until} $\lambda_{\rm REST} > 1.0$}
\item{\begin{itemize}
	\item{delete the simplices containing both $A$ and $A^\prime$}
	\item{replace $A$ by $A^\prime$ in all simplices surrounding $A$}
	\item{set the correct neighborship relations in these simplices}
	\item{update the simplices incident to $A^\prime$ with flips}
	\end{itemize}}
\end{enumerate}
The simplices containing both $A$ and $A^\prime$ will change their orientation
in the last step, since their volume vanishes when $A$ and $A^\prime$ merge.
However, since these simplices are deleted anyway, their orientation does not
need to be maintained within this last step.
The orientation of the simplices containing $A$ but not $A^\prime$ (described 
by $\lambda_{\rm REST}$) however, needs to be maintained, since these 
simplices will not be deleted afterwards.
Therefore, the quantity $\lambda_{\rm REST}$ should be the criterion for the 
last vertex step, whereas $\lambda$ accounts for the maximum length of the 
previous steps.

Another problem is posed by rounding errors in (\ref{Econdition}): If the
numerator becomes very small -- i.e.~if one has simplices with an extremely
small volume or very skinny simplices, then $\lambda$ may tend to assume
very small values. Rounding errors are then likely to happen as well.
This problem can be weakened by using exact arithmetics \cite{devillers2002a} when computing 
(\ref{Econdition}) or -- when working with random data -- by distributing the
data over a larger region of space (larger simplices). In our experiences with
triangulations of reasonable data such situations actually never occurred.

We have run several tests on the deletion algorithm by first triangulating
a number of points and then deleting all the points one by one. Before deleting a point
we collected all simplices not containing the point (roughly speaking: the triangulation
with a star-shaped hole) and checked for their existence in the resulting triangulation
after point deletion. In accordance with the theorem proved in \cite{shewchuk1998}
they were all recovered in the final triangulation -- which has also been compared with
one constructed by complete re-triangulation.

\section{Performance}\label{Sresults}

To test our implementation, we performed calculations on a $1.533$ GHz AMD 
Athlon processor with 1 GByte of RAM. The code has been compiled using the 
GNU g$++$ 3.3 compiler with compiler optimization set. The times were then obtained 
using the clock() command. The seed values of the random number generator 
have been determined using the system time. In all test runs, the data 
consisted of 64-bit double variables.

\subsection{Incremental Insertion Algorithm}\label{SSincinsalg}

The complexities of the walk algorithm and of the Green-Sibson 
algorithm have been extensively studied \cite{Edelsbrunner1996,devillers2001}. Here we 
have studied the computation time in dependence of the number of points to be 
triangulated. Test runs were performed for different configurations of points 
ranging from $10^3$ to $10^6$. 
To avoid the handling of vertex
rejection we sticked to the case of equally-weighted vertices (standard Voronoi tessellation).

\begin{figure}[ht]
  \centering {\includegraphics[height=8cm]{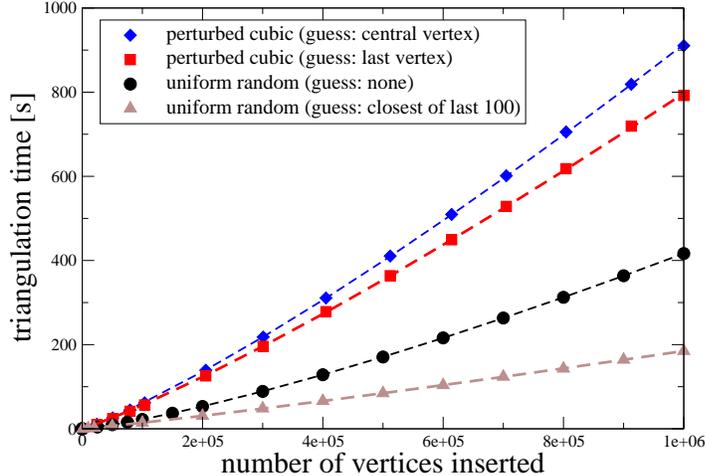}}
  \caption{\label{Finsertion_times}
	Times necessary for the tetrahedralization of different point numbers
	for different distributions. Cubic lattices are known to produce many 
	flat simplices (called slivers).
	In the case of the points distributed on (perturbed) lattices, the cost of the 
	simplex location can be reduced to constant by giving a good first 
	guess.
	In the case of randomly distributed points the walk in the triangulation can be considerably 
	shortened by choosing a better guess for a starting simplex.
	Dashed lines are fits with the expected overall algorithmic 
	complexities $\alpha_i N^{4/3} + \beta_i N$.}
\end{figure}

In a first series of runs, we considered a slightly perturbed cubic lattice 
with the average lattice constant $a=1.0$ (diamonds in figure \ref{Finsertion_times}).
As a starting simplex for the simplex walk we always took a simplex in the
center of the cubus. The expected algorithmic complexity of 
$\alpha_c N^{4/3} + \beta_c N$ is in complete agreement with the simulation.
In a second test run, we took the same lattice configuration but gave an imperfect guess
for the walk algorithm (squares in fig.~\ref{Finsertion_times}). 
This guess was the last simplex created and therefore good within the cubus 
and bad at the surface of the cubus. For large numbers of points -- where the ratio 
between vertices at the boundary of the cubus and the total number of vertices
in the cubus becomes small -- we find a linear behavior, as is
expected if the cost for simplex location becomes constant.
To test for robustness of our code we also fed an unperturbed cubic lattice (data not shown),
in this case one finds the worst-case quadratic scaling of 
$T_{\rm run}\approx 8.2 10^{-8} N^2$. However, in our calculations such extreme 
triangulations will never occur.
For comparison with a uniform random distribution we triangulated 
different numbers of points within the cubus 
$[-10.0, +10.0]\times[-10.0, +10.0]\times[-10.0, +10.0]$. A much better behavior of
the algorithmic complexity is found (spheres in fig.~\ref{Finsertion_times}).
Since for random data nothing is known about the final neighborships, no good guess 
can be given without some sort of preprocessing.
However, by choosing some simplex associated with the closest vertex 
(out of the last 100 inserted) in analogy to \cite{muecke1996,devroye1998}
one still finds a considerable gain in efficiency and a nearly linear scaling in the 
observed range (triangles in fig.~\ref{Finsertion_times}).
Furthermore, figure \ref{Finsertion_times} shows that the running times for random data
with our simple data structure are comparable with the more sophisticated
three-dimensional DCFL data structure \cite{Ferrez2001}, and other code 
\cite{muecke1998}, where the used algorithms code scale similarly on 
random data.

\begin{figure}[ht]
  \centering {\includegraphics[height=8cm]{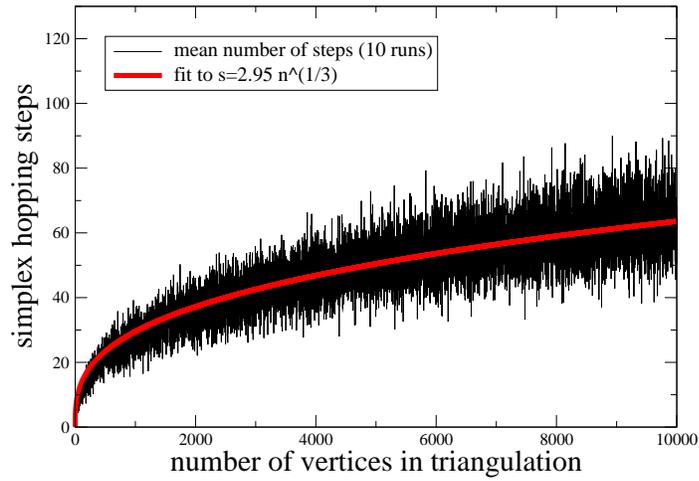}}
  \caption{\label{Fstep_number}
	The number of necessary steps starting from an arbitrary simplex in the
	triangulation towards another arbitrary simplex scales like $n^{1/3}$.
 	Shown is the mean out of ten runs.}
\end{figure}

It is evident that the incremental insertion of data points depends on
the choice of the initial simplex. Figure \ref{Fstep_number} shows the 
increase in the average number of steps necessary for the location of 
the simplex containing a point using the visibility walk. 
The expected average $n^{1/3}$ relation is found.

\subsection{Incremental Deletion Algorithm}\label{SSincdeletalg}

\begin{figure}[ht]
  \centering {\includegraphics[height=8cm]{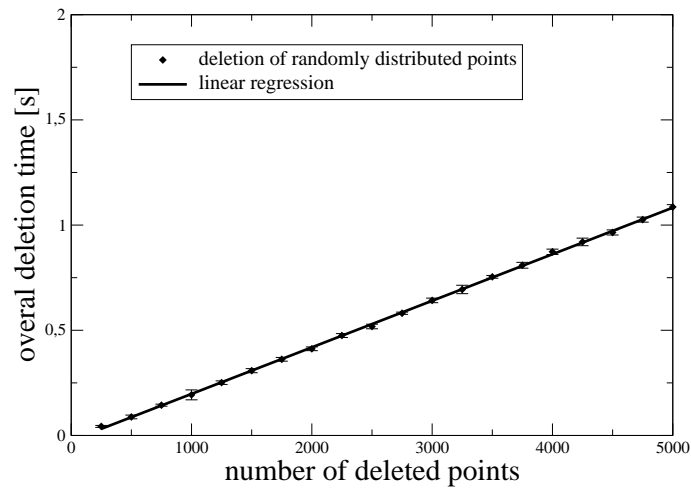}}
  \caption{\label{Fdeletion_time}
	Calculation times necessary for the deletion of different
	numbers of points.}
\end{figure}

In simulation of growth models it will often be necessary to delete 
vertices from a Delaunay triangulation. 
It turns out, that vertex deletion is more efficient than vertex insertion, since there
is no cost associated with simplex location, as the vertices provide $\order{1}$ access
to the incident simplices.
Furthermore, one would expect the average algorithmic
complexity of vertex deletion to be constant (i.e.~not to depend on the total
number of points). In this experiment we have first created a Delaunay 
triangulation and deleted it afterwards by removing point by point. Again, the
mean out of ten test runs has been calculated. Figure \ref{Fdeletion_time} 
gives an impression of the expected linear relationship.

\subsection{Restoring the Delaunay property}\label{SSrestore}

A simulation hosting kinetic vertices will be especially sensitive on the
cost of checking all simplices for Delaunay invalidity and restoring the 
Delaunay property. 
We have triangulated $N\in [ 10^4,10^6 ]$ random points in a vectorizable random lattice \cite{lauritsen1994}
(equivalent to a strongly perturbed cubic lattice) with the additional
condition of a minimum distance of 0.1 between all the points. This seemed more realistic to
us, since in proximity structures there will always be some minimum distance defined by the
object sizes.
Afterwards all points are moved by a random small step towards a hypothetical new position
$\vec X \to \vec X + [RND(-m, m), RND(-m, m), RND(-m, m)]$ with $m$ chosen constant.
These hypothetical new positions are put on a list, which is being iterated:
Possible steps that do not invalidate the triangulation are performed instantly, the others 
are divided in smaller sub-steps where the criterion for the maximum step size is the same as 
that in section \ref{SSdeletion}.
Then the Delaunay criterion is restored (with $2\to 3$ or $3\to 2$ flips).
The algorithm terminates when all new positions have been reached. 

In practice, this
procedure would correspond to one timestep of the application and in the ideal case (where the
step sizes are small enough) just one iteration should suffice.
However, the additional cost of performing a step size check before actually moving the vertices
produces only a factor of roughly 2 in the restoration time and should therefore be preferred to
increase the robustness of the algorithm considerably.

\begin{figure}[ht]
  \centering {\includegraphics[height=8cm]{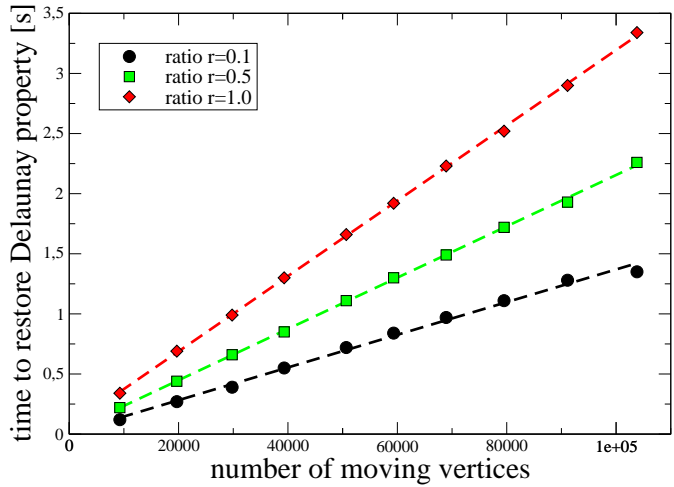}}
  \caption{\label{Fupdate_times}
	Shown are the calculation times necessary for the restoration of the
	Delaunay criterion after vertex movement for different ratios $r=m/d_{\rm min}$ 
	(step size over the minimum distance) versus the number of points.
	The expected linear relation is found with slopes increasing with the step size.
	In the ideal case, this update method is about 20 times as fast as computing 
	a new triangulation.}
\end{figure}

As expected, the complexity behaves linear in the number of points, see e.g.~figure \ref{Fupdate_times}.
By looking at figure \ref{Fupdate_times} one finds that restoring the Delaunay criterion in a 
slightly perturbed Delaunay triangulation being about 20 times as fast as recomputing the whole 
triangulation as long as the vertex displacements are small compared to the average vertex distance.
Locally updating a triangulation is also faster than using a combination of delete and insert 
operations \cite{fabritiis2003,lauritsen1994}, since much less simplices have to be flipped.

\subsection{Mixed algorithms}\label{SSmixed}

To check whether a simulation can cope with a varying number of kinetic
vertices, we combined the algorithms on vertex insertion, vertex deletion and
vertex movement. For different numbers of uniformly distributed vertices 100 
time steps have been performed. In each time step, with probability 
$p=0.5$ an arbitrary vertex was deleted from the triangulation and with
probability $p=0.5$ a random vertex was inserted. Afterwards all the vertices
were moved by a small deviation followed by the restoration of the Delaunay 
criterion. Therefore, if in average a constant number of vertices are deleted or inserted
per timestep, we can expect an overall linear behavior as in
table \ref{Tmixed}.

\begin{table}[ht]\label{Tmixed}
\caption{Code performance for different numbers of vertices. In every run, 100
	timesteps have been performed. In each timestep, with probability 
	$p=0.5$ either an old vertex was deleted or a new vertex was inserted
	into the triangulation (second and third columns). Then all the 
	vertices were moved by a small amount and the flips necessary to 
	restore the Delaunay criterion have been counted -- the fourth column
	does not include the flips necessary for the deletion process.
	In the last column, the calculation time per timestep is given.}
\begin{tabular}{c|c|c|c|c}
points & deletions (total) & insertions (total) & flips (total)& 
	one timestep [s]\\
\hline
20000  & 59 & 41 &  426 & 0.26\\
40000  & 51 & 49 & 1098 & 0.54\\
60000  & 52 & 48 & 2067 & 0.84\\
80000  & 46 & 54 & 2749 & 1.15\\
100000 & 42 & 58 & 3521 & 1.47\\
120000 & 47 & 53 & 5154 & 1.81\\
140000 & 62 & 38 & 6297 & 2.14\\
160000 & 56 & 44 & 7207 & 2.49\\
180000 & 50 & 50 & 7918 & 2.84\\
200000 & 49 & 51 & 9766 & 3.21\\
\end{tabular}
\end{table}

\section{Summary}\label{Ssummary}

In this article we have shown that it is possible to construct 
fully dynamic and kinetic three-dimensional Delaunay triangulations by using a
very simple data structure.
This data structure is obtained by adding neighborship entries to every simplex and by 
storing the tetrahedra within a list. The performance of our data structure is
comparable to that of more sophisticated kinetic data structures \cite{Ferrez2001}, 
which may pose an advantage for parallelization. 

We have proposed a new incremental method of vertex deletion which uses a maximum step size
criterion. This criterion also solves the problem of maintaining a valid three-dimensional
Delaunay triangulation during vertex movement.

Note that the code allows to be generalized towards power-weighted Delaunay 
triangulations in a mostly straightforward way by replacing the normal circumsphere 
criterion by its weighted counterpart.
In addition, the code provides functionality to compute volumes and 
contact surfaces of the associated Voronoi cells which are of importance in
some simulations of interacting particle systems.
The resulting tessellation of space in Voronoi/Laguerre cells can be used to
model growth/shrinking processes or for the numerical solution of differential
equations on irregular grids. 
This implementation of a fully dynamic and kinetic Delaunay triangulation
thus makes our code suitable for the simulation of dynamically interacting 
complex systems with variable particle numbers as e.g.~cell tissues.

\section{Acknowledgments}\label{Sacknowledge}
G.~S.~is indebted to T.~Beyer for discussing many aspects of the algorithms
and for testing the code and to U.~Brehm for discussing many theoretical 
problems. G.~S.~has been financially supported by the SMWK.

\bibliography{phdthesis}
\bibliographystyle{elsart-num} 

\end{document}